\documentclass[preprint,footinbib]{revtex4-1}
\usepackage{amsmath}
\usepackage{amsfonts}
\usepackage{graphicx}
\usepackage{bbold}
\usepackage{csquotes}
\usepackage[font=scriptsize,labelfont=bf]{caption}
\usepackage{amssymb}

\begin{document}

\title{Localised solution of stochastic Schr\"odinger equation describing simultaneous continuous measurement of position and momentum}
\date{\today}
\author{Daniel J.~Bedingham}
\email{daniel.bedingham@rhul.ac.uk}
\affiliation{Department of Physics, Royal Holloway, University of London, Egham, TW20 0EX, United Kingdom.}

\begin{abstract}
A stochastic Schr\"odinger equation is presented to describe simultaneous continuous measurement of the position and momentum of  a non-relativistic particle. The equation is solved to yield a state localised in position and momentum contingent on the uncertainty principle. The state is understood as that to which the particle tends after a period of continuous measurement. The solution takes the form of a wave packet of fixed Gaussian shape whose centre undergoes random motion in phase space. It is shown to be locally stable.
\end{abstract}
\maketitle


The Heisenberg uncertainty principle states that there is a limit to the precision with which the position and momentum of a particle can be jointly known. This implies that it is not possible to do precise simultaneous measurements of position and momentum of a particle. However, if the measurements are imprecise there is nothing in principle to prevent them from being carried out. Furthermore, by performing a sequence of imprecise measurements we would expect that the particle would tend towards a state of localised position and momentum subject to uncertainty principle constraints.

The idea of simultaneous imprecise measurement of position and momentum has been explored theoretically in Refs \cite{REF2, REF12} and an analogous situation in a qubit system of simultaneously measuring non-commuting observables has been performed experimentally \cite{REF1}. Several others have explored features of simultaneous measurements of non-commuting observables, e.g.~\cite{REF6,REF4,REF5,REF10,REF9,REF7}.

Here we consider a model of continuous simultaneous measurement of both position and momentum described by a stochastic Schr\"odinger equation. We will give an exact solution for the equation in the form of a localised wave function of fixed shape centred about a stochastically varying point in phase space. We will show that our solution is locally stable, so that, in the neighbourhood of the solution, the stochastic dynamics tend to drive the particle to this state of localisation. Our method of solution borrows from Di\'osi \cite{DIO2} who considered an equation describing continuous position measurement of a non-relativistic particle (a model sometimes known as QMUPL, quantum mechanics with universal position localisation \cite{DIO1}).

Our model can be interpreted as a collapse model \cite{GRW, CSL1, CSL2} (for reviews see \cite{DRM,Pearlebook}) in which both position and momentum undergo spontaneous localisation. So far there has been little interest in collapse models involving non-commuting collapse generating operators. This is perhaps due to the difficulty of analysing or finding solutions in such cases. 

We start with the stochastic Schr\"odinger equation
\begin{align}\label{SSE}
d|\psi\rangle = \bigg\{-\frac{ip^2}{2m}dt - \frac{1}{4}\gamma\left(q - \langle q\rangle \right)^2 dt
&- \frac{1}{4}\gamma'\left(p - \langle p\rangle \right)^2dt  \nonumber \\
&+ \left(q - \langle q\rangle \right)d\xi + \left(p - \langle p\rangle \right)d\xi' \bigg\}|\psi\rangle,
\end{align}
where quantum expectations are denoted by
\begin{align}\label{QE}
\langle \cdot \rangle = \langle \psi | \cdot  | \psi \rangle.
\end{align}
The Wiener processes $\xi$ and $\xi'$ satisfy
\begin{align}
\mathbb{E}[d\xi] = \mathbb{E}[d\xi'] = 0,
\end{align}
where $\mathbb{E}[\cdot]$ denotes stochastic expectation, and 
\begin{align}
d\xi d\xi = \frac{1}{2}\gamma dt\; ; \; d\xi' d\xi' = \frac{1}{2}\gamma' dt \; ; \; d\xi' d\xi = 0.
\end{align}
Equation (\ref{SSE}) describes the continuous simultaneous measurement of position and momentum of a non-relativistic quantum particle (see, e.g.~\cite{REF13}). The parameters $\gamma$ and $\gamma'$ determine the precision of the position and momentum measurements respectively. In the context of collapse models these parameters are known as the collapse strengths.

Under the dynamics described by equation (\ref{SSE}), the process followed by the quantum expectation of an operator $O$ is found, using the usual rules of It\^o calculus, to be 
\begin{align}\label{<O>}
d\langle O \rangle = \langle i\left[p^2/2m,O\right]\rangle dt  &- \frac{1}{4}\gamma \langle \left[q,\left[q,O\right]\right] \rangle dt 
- \frac{1}{4}\gamma' \langle \left[p,\left[p,O\right]\right] \rangle dt \nonumber \\
&+\langle \{O,(q-\langle q\rangle) \}\rangle d\xi +\langle \{O,(p-\langle p\rangle) \}\rangle d\xi'.
\end{align}
In particular we can use this formula to work out the processes followed by the quantum expectations of particle position $\langle q\rangle$ and particle momentum $\langle p\rangle$:
\begin{align} \label{<q>}
d\langle q\rangle &= \frac{\langle p\rangle}{m} + 2\sigma^2d\xi + 2R d\xi', \\
d\langle p\rangle & = 2Rd\xi + 2\sigma'^2d\xi',  \label{<p>}
\end{align}
where we have defined
\begin{align}
\sigma^2\, &= \langle\left(q - \langle q\rangle \right)^2 \rangle, \\
\sigma'^2 &= \langle\left(p - \langle p\rangle \right)^2 \rangle,\\
R \;& = \frac{1}{2}\langle \{ \left(q - \langle q\rangle \right),  \left(p - \langle p\rangle \right)   \}\rangle.
\end{align}
These are interpreted respectively as the quantum variance in position, the quantum variance in momentum, and the quantum covariance between position and momentum. 

In order to solve equation (\ref{SSE}) we follow the method of Di\'osi \cite{DIO2} by performing a phase space shift to the comoving frame
\begin{align} \label{comovestate}
|\tilde{\psi}\rangle = e^{-i\langle p \rangle q}e^{i \langle q \rangle p} |\psi\rangle.
\end{align}
This transformation means that quantum expectations of position and momentum calculated using the state $|\tilde{\psi}\rangle$ are zero. Note that in what follows, $\langle\cdot \rangle $ continues to denote quantum expectation using the state $|\psi\rangle$. Also note that for quantum expectations $\sigma$, $\sigma'$ and $R$, there is no change upon replacing $|\psi\rangle \rightarrow|\tilde{\psi}\rangle $.

In order to find the process for the state $|\tilde{\psi}\rangle $ we need
\begin{align}
d[e^{-i\langle p \rangle q}] &= \left\{
-2iRqd\xi -\gamma R^2 q^2  dt - 2i\sigma'^2 q d\xi' - \gamma'\sigma'^4 q^2dt
\right\}e^{-i\langle p \rangle q}, \\
d[e^{i \langle q \rangle p}] &= \left\{
\frac{i\langle p\rangle p}{m}dt + 2i\sigma^2 p d\xi - \gamma \sigma^4 p^2 dt + 2iRpd\xi' - \gamma'R^2p^2dt
\right\}e^{i \langle q \rangle p}, 
\end{align}
from which, using the multivariate It\^o lemma on (\ref{comovestate}) we find
\begin{align}\label{psitildeproc}
d|\tilde{\psi}\rangle =& \left(q-2iRq + 2i\sigma^2p \right) d\xi |\tilde{\psi}\rangle +2i\sigma ^2\langle p \rangle d\xi|\tilde{\psi}\rangle \nonumber\\
&+  \left(p-2i\sigma'^2q + 2iR p\right) d\xi' |\tilde{\psi}\rangle +2i R\langle p \rangle d\xi'|\tilde{\psi}\rangle \nonumber\\
& +\frac{1}{4}\gamma \left(q-2iRq + 2i\sigma^2 p \right)^2dt |\tilde{\psi}\rangle
+\frac{1}{4}\gamma' \left(p-2i\sigma'^2 q + 2iR p\right)^2dt |\tilde{\psi}\rangle \nonumber\\
&+\left[ \frac{-ip^2}{2m} - \frac{1}{2}\gamma(q^2-\sigma^2) - \frac{1}{2}\gamma'(p^2 - \sigma'^2) 
+ i\gamma \sigma^2 R + i\gamma' \sigma'^2 R 
\right]dt |\tilde{\psi}\rangle \nonumber\\
&+\frac{i\langle p\rangle ^2}{2m} dt|\tilde{\psi}\rangle
+\left(-2\gamma\sigma^4 \langle p\rangle p  - \gamma \sigma^4 \langle p\rangle^2
+2\gamma \sigma^2 R  \langle p\rangle q + i \gamma  \sigma^2 \langle p\rangle q
\right)dt|\tilde{\psi}\rangle \nonumber\\
&+\left(-2\gamma' R^2 \langle p\rangle p  - \gamma'R^2\langle p\rangle^2 + 2\gamma' \sigma'^2 R \langle p\rangle q
+ i \gamma'  R \langle p\rangle p \right)dt|\tilde{\psi}\rangle.
\end{align}
Notice that even though we have shifted to the comoving frame, the process for the state $|\tilde{\psi}\rangle $ still depends on $\langle q\rangle$ and $\langle p\rangle$. In order to simplify and close the system we choose the origin of the phase space coordinate system such that, momentarily, $\langle q\rangle = \langle p\rangle = 0$. At this point in time $|\psi\rangle = |\tilde\psi\rangle$ [see equation (\ref{comovestate})], and (\ref{psitildeproc}) simplifies to 
\begin{align} \label{psitilde}
d|\tilde{\psi}\rangle =& \left(q-2iRq + 2i\sigma^2 p \right) d\xi |\tilde{\psi}\rangle + 
\left(p-2i\sigma'^2q + 2iRp \right) d\xi' |\tilde{\psi}\rangle \nonumber\\
& +\frac{1}{4}\gamma \left(q-2iRq + 2i\sigma^2p \right)^2dt |\tilde{\psi}\rangle
+\frac{1}{4}\gamma' \left(p-2i\sigma'^2q + 2iRp \right)^2dt |\tilde{\psi}\rangle \nonumber\\
&+\left[ \frac{-ip^2}{2m} - \frac{1}{2}\gamma(q^2-\sigma^2) - \frac{1}{2}\gamma'(p^2 - \sigma'^2) 
+ i\gamma \sigma^2 R+ i\gamma' \sigma'^2 R 
\right]dt |\tilde{\psi}\rangle.
\end{align}
This is a closed system since $\sigma^2, \sigma'^2, R$ can all be defined using $|\tilde\psi\rangle$. 

In order to solve equation (\ref{psitilde}) we assume the stationary ansatz  
\begin{align}
\langle q |\tilde{\psi}\rangle \rightarrow \langle q |\tilde{\psi}_{\infty}\rangle = \tilde{\psi}_{\infty}(q) e^{-iEt},
\end{align}
where $E$ is some constant.
Inserting this ansatz into (\ref{psitilde}), and equating stochastic terms we find
\begin{align}
\langle q |\left(q-2iRq + 2i\sigma^2 p\right)|\tilde{\psi}_\infty\rangle &= 0, \label{eq1} \\
\langle q |\left(p-2i\sigma'^2q + 2iRp \right)|\tilde{\psi}_\infty\rangle &= 0,\label{eq2}
\end{align}
and equating drift terms we find
\begin{align}
\langle q |\left[ \frac{-ip^2}{2m} - \frac{1}{2}\gamma(q^2-\sigma^2) - \frac{1}{2}\gamma'(p^2 - \sigma'^2) 
+ i\gamma \sigma^2 R + i \gamma'  \sigma'^2 R  \right]|\tilde{\psi}_\infty\rangle = -iE\langle q|\tilde{\psi}_\infty\rangle.\label{eq3}
\end{align}

Equations (\ref{eq1}), (\ref{eq2}) and (\ref{eq3}) can be expressed as differential equations:
\begin{align}\label{eq4}
\frac{\partial}{\partial q}\tilde{\psi}_\infty(q) =\frac{-(1-2iR)}{2\sigma^2} q \tilde{\psi}_\infty(q),
\end{align}
\begin{align}\label{eq5}
\frac{\partial}{\partial q}\tilde{\psi}_\infty(q) =\frac{-2\sigma'^2}{(1+2iR)} q \tilde{\psi}_\infty(q),
\end{align}
\begin{align}\label{eq6}
\left[\left(\frac{-1}{2m} + \frac{1}{2}\gamma' i \right)\frac{\partial^2}{\partial q^2}
-\frac{1}{2}i \gamma  (q^2 - \sigma ^2) + \frac{1}{2}i \gamma'  \sigma'^2
\right] \tilde{\psi}_\infty(q) = E' \tilde{\psi}_\infty(q),
\end{align}
where
\begin{align}
E' = E + \gamma \sigma^2 R + \gamma'\sigma'^2 R. 
\end{align}

These equations are straightforward to solve. We begin by solving (\ref{eq4}):
\begin{align}\label{soln1}
\tilde{\psi}_\infty(q) = \frac{1}{\sqrt[4]{2\pi\sigma^2}} 
\exp\left\{  \frac{-(1-2iR)}{4\sigma^2} q^2 \right\}.
\end{align}
This gives the shape of the stationary state as a Gaussian. The constant factor is chosen to normalise the state.

Note that equation (\ref{eq5}) results in the same solution provided that
\begin{align}
\frac{(1-2iR)}{2\sigma^2} = \frac{2\sigma'^2}{(1+2iR)},
\end{align}
or
\begin{align}\label{constraint}
1 + 4R^2 = 4\sigma^2\sigma'^2.
\end{align}
At this stage it can be confirmed that, e.g.,
\begin{align}
 \langle \tilde\psi_\infty |q^2| \tilde\psi_\infty \rangle = \int dq \; q^2 |\tilde{\psi}_\infty(q) |^2 = \sigma^2,
\end{align}
and similarly $ \langle \tilde\psi_\infty |p^2| \tilde\psi_\infty \rangle = \sigma'^2$, and $ \langle \tilde\psi_\infty |qp + pq| \tilde\psi_\infty \rangle = 2R$. Thus $\sigma^2$, $\sigma'^2$, and $R$ respect their definitions as quantum expectations.

Inserting (\ref{soln1}) into equation (\ref{eq6}), separating coefficients of different powers of $q$, and taking real and imaginary parts we find
\begin{align}
0 &= \frac{1}{4m\sigma^2} -\frac{\gamma'R}{2\sigma^2} - E' = 0, \\
0 &= -\frac{R}{2m\sigma^2} -\frac{\gamma'}{4\sigma^2} + \frac{1}{2}\gamma\sigma^2 + \frac{1}{2}\gamma'\sigma'^2, 
\label{28}\\
0 &= -\frac{(1-4R^2)}{4m} + \gamma'R, 
\label{29}\\
0 &= \frac{R}{2m\sigma^4} + \frac{\gamma'(1-4R^2)}{8\sigma^4} -\frac{1}{2}\gamma.
\label{30}
\end{align}
This is a set of 4 simultaneous equations for $\sigma^2$, $\sigma'^2$, $R$, and $E'$. They are easily solved to find
\begin{align}
\sigma^2 = \sigma^2_\infty = \sqrt{\frac{1}{2m\gamma}} \sqrt{m^2\gamma'^2 +1}\left(\sqrt{m^2\gamma'^2 +1} - m\gamma' \right)^{1/2},
\end{align}
\begin{align}
\sigma'^2 = \sigma'^2_\infty = \sqrt{\frac{\gamma m}{2}}\left(\sqrt{m^2\gamma'^2 +1} - m\gamma' \right)^{1/2},
\end{align}
\begin{align}\label{Rinfty}
R = R_{\infty} = \frac{1}{2}\left(\sqrt{m^2\gamma'^2 +1} - m\gamma' \right),
\end{align}
and
\begin{align}
E' = \frac{1}{2\sigma_\infty^2}\left(\frac{1}{2m}-\gamma'R_\infty\right). 
\end{align}
We note that the condition (\ref{constraint}) is satisfied by these forms. The solution satisfies the uncertainty principle since, from (\ref{constraint}),
\begin{align}
\sigma_{\infty}\sigma'_{\infty} = \sqrt{\frac{1}{4} + R_\infty^2},
\end{align}
and therefore, $\sigma_{\infty}\sigma'_{\infty} \geq \frac{1}{2}$.

Having found the form of the solution for $|\tilde\psi\rangle$ with $\langle q\rangle = \langle p\rangle = 0$, we can undo our self-imposed constraints. First, to solve for $|\tilde\psi\rangle$ with general $\langle q\rangle$ and $\langle p\rangle$ we return to equation (\ref{psitildeproc}) and write the solution in the form 
\begin{align}
\langle q |\tilde{\psi}_{\infty}\rangle  = \tilde\psi_{\infty} (q) f_t
\end{align}
where $\tilde\psi_{\infty} (q)$ is given above and $f_t$ is some stochastic process independent of $q$. Here we are assuming that due to Galilean invariance, the effect of shifting the phase space origin should have no effect on the shape of the wave function $ \tilde\psi_{\infty} (q)$ and should only contribute a phase factor $f_t$.

Inserting the solution in this form we find that the equation is solved by
\begin{align}
f_t = e^{2i\sigma^2\int^t\langle p\rangle d\xi}e^{2iR\int^t \langle p\rangle d\xi'}e^{\frac{i}{2m}\int^t \langle p\rangle^2 dt} e^{-iEt}.
\end{align}

Finally we undo the phase space shift to the state
\begin{align}
\langle q|\psi_{\infty}\rangle = \langle q |e^{-i \langle q \rangle p}e^{i\langle p \rangle q}|\tilde{\psi}_\infty\rangle.
\end{align}
This results in the full solution to equation (\ref{SSE})
\begin{align}\label{soln}
\langle q |\psi_\infty\rangle = \frac{1}{\sqrt[4]{2\pi\sigma^2}} e^{i\langle p\rangle q}
e^{\frac{-(1-2iR)}{4\sigma^2} (q-\langle q\rangle)^2}
e^{2i\sigma^2\int^t\langle p\rangle d\xi}e^{2iR\int^t \langle p\rangle d\xi'}e^{\frac{i}{2m}\int^t \langle p\rangle^2 dt} e^{-iEt}
e^{-i\langle p\rangle \langle q\rangle},
\end{align}
where
\begin{align}
E = \frac{1}{2\sigma^2}\left(\frac{1}{2m}-\gamma'R\right) - \gamma \sigma^2 R - \gamma' \sigma'^2 R,
\end{align}
and where $\sigma^2,\sigma'^2, R$, are given by $\sigma^2_\infty,\sigma'^2_\infty,R_\infty$ shown above. This is our main result. The $q$-dependent terms in (\ref{soln}) describe a solution which maintains a fixed Gaussian shape centred about $\langle q\rangle$,$\langle p\rangle$ in phase space. Expected position and momentum vary stochastically according to equations (\ref{<q>}) and (\ref{<p>}) describing a wave packet which randomly moves around in phase space. The overall phase factor also varies stochastically since it depends on the quantum expectations of position and momentum along with stochastic variables $\int^t \langle p\rangle d\xi$ and $\int^t \langle p\rangle d\xi'$.
 
We can confirm (\ref{soln}) by inserting into equation (\ref{SSE}) and comparing with the process for $\langle q |\psi_\infty\rangle$ obtained by application of the multivariate It\^o lemma to (\ref{soln}) with stochastic variables $\langle p\rangle $, $\langle q \rangle$, $\int^t \langle p\rangle d\xi$ and $\int^t \langle p\rangle d\xi'$.

Having found a localised solution of equation (\ref{SSE}) it remains to show that this solution is stable. To demonstrate this we will prove local stability by showing that the optimal values $\sigma^2_\infty$, $\sigma'^2_\infty$, and $R_\infty$ constitute a dynamical fixed point, whereby, for any small deviations away from these values, there is a dynamical drive to return to them. We do not prove global stability although it would seem a reasonable conjecture. 

Using (\ref{<O>}) we can calculate the processes followed by $\langle q^2\rangle$, $\langle p^2\rangle$ and $\langle \{q,p\}\rangle$. Then using (\ref{<q>}), (\ref{<p>}) and the It\^o formula we can calculate the processes for $\sigma^2$, $\sigma'^2$, and $R$. Finally taking the stochastic expectation of the differentials we find
\begin{align}\label{41}
\mathbb{E}\left[d\sigma^2\,\right] &=
\left(\frac{2R}{m}+ \frac{\gamma'}{2} - 2\gamma\sigma^4 dt - 2\gamma'R^2 \right)dt,  \nonumber\\
\mathbb{E}\left[d\sigma'^2\right] &= 
\left(\frac{\gamma}{2} - 2\gamma R^2- 2\gamma'\sigma'^4\right) dt, \nonumber\\
\mathbb{E}\left[\;dR\;\right] &= \left(\frac{\sigma'^2}{m}-2\gamma \sigma^2 R- 2\gamma'\sigma'^2R \right)dt.
\end{align}
These are interpreted as the expected incremental changes in $\sigma^2$, $\sigma'^2$, and $R$ during a time increment $dt$. It can be shown that when the three expressions in curved brackets are each set equal to zero, then equations (\ref{constraint}), (\ref{28}), (\ref{29}), and (\ref{30}) are all satisfied. This shows that when the state reaches a point where $\sigma^2$, $\sigma'^2$, and $R$ remain constant in stochastic expectation, then the values for these variables are $\sigma^2_\infty$, $\sigma'^2_\infty$, and $R_\infty$. Until this point these variables will drift.

Let us write 
\begin{align}
\sigma^2\, &= \sigma^2_\infty + x_1\nonumber\\
\sigma'^2 &= \sigma'^2_\infty + x_2\nonumber\\
R\; &= R_\infty + x_3
\end{align}
so that ${\bf x} = (x_1, x_2, x_2)$ represents small deviations of $\sigma^2$, $\sigma'^2$, and $R$ from their optimal values. Working to linear order in ${\bf x}$ we can then express (\ref{41}) as
\begin{align}
\mathbb{E}[{d\bf x}] = A{\bf x}dt
\end{align}
where
\begin{align}
A = 
\begin{bmatrix}
-4\gamma\sigma^2_\infty & 0 & \left(\frac{2}{m}-4\gamma'R_\infty\right) \\
0 & -4\gamma'\sigma'^2_\infty & -4\gamma R_\infty \\
-2\gamma R_\infty & \left(\frac{1}{m}-2\gamma'R_\infty\right) & (-2\gamma\sigma^2_\infty - 2\gamma'\sigma'^2_\infty)
\end{bmatrix}.
\end{align}

The eigenvalues of $A$ are found to be
\begin{align}
\lambda = -(2\gamma \sigma^2_\infty + 2\gamma'\sigma'^2_\infty);\;\; -(2\gamma \sigma^2_\infty + 2\gamma'\sigma'^2_\infty) 
\pm 2i\sqrt{(\gamma\sigma^2_\infty - \gamma'\sigma'^2_\infty)^2 + 4\gamma\gamma'R^2_\infty}.
\end{align}
Since all have negative real components then the fixed point ${\bf x} = {\bf 0}$ is a stable attractor. In other words there is an expected drift towards the fixed point. Values of $\sigma^2$, $\sigma'^2$, and $R$ deviating by a small amount from $\sigma^2_\infty$, $\sigma'^2_\infty$, and $R_\infty$ will tend towards these values in stochastic expectation.

In summary we have shown that for the stochastic Schr\"odinger equation (\ref{SSE}) which describes simultaneous continuous measurement of both position and momentum of a non-relativistic quantum particle, there is a localised solution given by (\ref{soln}). The solution takes the form of a Gaussian wave packet of fixed shape whose centre undergoes random motion in phase space, and whose overall phase also changes randomly. We have shown that the stochastic dynamics drive the particle towards this state of localisation indicating that the solution is at least locally stable.



\begin{thebibliography}{99}

\bibitem{REF2}
E.~Arthurs \& J.~L.~Kelly, Bell Syst.~Tech.~J.~{\bf 44}, 725 (1965).

\bibitem{REF12}
C.~Y.~She \& H.~Heffner, Phys.~Rev.~{\bf 152}, 1103 (1966).

\bibitem{REF1}
S.~Hacohen-Gourgy, L.~S.~Martin, E.~Flurin, V.~V.~Ramasesh, K.~ B.~Whaley, \& I.~Siddiqi, Nature {\bf 538}, 491 (2016).

\bibitem{REF6}
M.~Perarnau-Llobet \& T.~M.~Nieuwenhuizen, Phys.~Rev.~A{\bf 95}, 052129 (2017).

\bibitem{REF4}
A.~Chantasri, J.~Atalaya, S.~Hacohen-Gourgy, L.~S.~Martin, I.~Siddiqi, \& A.~N.~Jordan, Phys.~Rev.~A{\bf 97}, 012118 (2018). 

\bibitem{REF5}
M.~A.~Ochoa, W.~Belzig, \& A.~Nitzan, Sci.~Rep.~{\bf 8}, 15781 (2018).

\bibitem{REF10}
C.~Jiang \& G.~Watanabe, Phys.~Rev.~A{\bf 102}, 062216 (2020)

\bibitem{REF9} 
C.~Jiang \& G.~Watanabe, Phys.~Rev.~A{\bf 105}, 022613 (2022)

\bibitem{REF7}
S.~M.~Walls \& I.~J.~Ford, Phys.~Rev.~A{\bf 110}, 032432 (2024).

\bibitem{DIO2} L.~Di\'osi, Phys.~Lett.~A{\bf 132}, 233 (1988).

\bibitem{DIO1} L.~Di\'osi, Phys.~Rev.~A{\bf 40}, 1165 (1989).

\bibitem{GRW}G.C.~Ghirardi, A.~Rimini \& T.~Weber, Phys.~Rev.~D{\bf 34}, 470 (1986);
Phys.~Rev.~D{\bf 36}, 3287 (1987); Found.~Phys.~{\bf 18}, 1 (1988).

\bibitem{CSL1} P.~Pearle, Phys.~Rev.~A {\bf 39}, 2277 (1989).

\bibitem{CSL2} G.C.~Ghirardi, P.~Pearle \& A.~Rimini, Phys.~Rev.~A{\bf 42}, 78 (1990). 

\bibitem{DRM} A.~Bassi \& G.C.~Ghirardi, Phys.~Rept.~{\bf 379}, 257 (2003).

\bibitem{Pearlebook} P.~Pearle, \textit{Introduction to Dynamical Wave Function Collapse} (OUP, Oxford, 2024).

\bibitem{REF13} K.~Jacobs \& D.~A.~Steck, Contemp.~Phys.~{\bf 47}, 279 (2006).




\end{thebibliography}
\end{document}